\documentclass[final,3p,times,twocolumn]{elsarticle}
\usepackage{graphicx}
\usepackage{amssymb}
\usepackage{amsmath}
\usepackage{hyperref}
\usepackage{color}
\usepackage{dsserif}
\usepackage[normalem]{ulem}
\journal{Physics Letters A{ \color{red} {\mbox(Accepted)}}}
\DeclareUnicodeCharacter{2212}{-}

\usepackage{hyperref}
\usepackage[T1]{fontenc} 

\begin{document}

\title{Supersolid-like solitons in two-dimensional nonmagnetic spin-orbit coupled  spin-1 and spin-2 condensates}

\author[fac]{Pardeep Kaur}
\ead{2018phz0004@iitrpr.ac.in}
\author[fac]{Sandeep Gautam}
\ead{sandeep@iitrpr.ac.in}
\cortext[author]{Corresponding author.}
 \address[fac]{Department of Physics, Indian Institute of Technology Ropar, Rupnagar-140001, Punjab, India}
\author[int]{S. K. Adhikari\corref{author}}
\ead{sk.adhikari@unesp.br}
\ead[url]{professores.ift.unesp.br/sk.adhikari/}
 \address[int]{Instituto de F\'{\i}sica Te\'orica, UNESP - Universidade Estadual Paulista, 01.140-070 S\~ao Paulo, S\~ao Paulo, Brazil}
      

\date{\today}

\begin{abstract}

We demonstrate  spontaneous generation of spatially-periodic supersolid-like  super-lattice and stripe solitons in Rashba spin-orbit (SO) coupled spin-1 and spin-2 quasi-two-dimensional nonmagnetic Bose-Einstein condensates (BECs). The solitons in  a weakly SO-coupled spin-1 BEC are circularly-symmetric of $(−1, 0, +1)$ and $(0, +1, +2)$ types and have inherent vorticity; the numbers in the parentheses are the winding numbers in  hyper-spin components $+1, 0, −1$, respectively.  The circularly-symmetric  solitons in an SO-coupled spin-2 BEC are of types $(−2, −1, 0, +1, +2)$ and $(−1, 0, +1, +2, +3)$ with the former being the ground state, where the winding numbers correspond to spin components 
$+2, +1, 0, −1, −2$, respectively. For stronger SO-coupling strengths, these solitons acquire a multiring structure while preserving the winding numbers. Quasi-degenerate stripe and super-lattice solitons, besides a circularly-asymmetric soliton, also emerge as excited stationary states for stronger SO-coupling strengths in spin-1 and spin-2 BECs.

\end{abstract}

\maketitle
 
\section{Introduction}

A quasi-one-dimensional (quasi-1D) and quasi-two-dimensional (quasi-2D) spin-orbit-coupled (SO-coupled) 
spin-1/2, spin-1, or spin-2 trapped or trapless Bose-Einstein condensates (BECs) can produce spatially-periodic
structures in component or total density 
\cite{Ketterle,Putra,Li, SS-spin-1/2,SS-SPIN1,SS-trapped spin-1, SS-spin-2}. 
A spin-1 or spin-2 spinor BEC may have distinct magnetic phases, like, ferromagnetic, anti-ferromagnetic, cyclic,
etc. \cite{review-spinor-cond}. {Distinct periodic structures are obtained in different magnetic phases of SO-coupled
spin-1 or spin-2 BECs \cite{SS-SPIN1,SS-trapped spin-1,SS-spin-2}. Hence, the spatially-periodic structures are
generally thought to be associated with these magnetic phases. In view of this, it is interesting to see if these 
periodic structures survive in the SO-coupled nonmagnetic spin-1 and spin-2 BECs.} In spin-1 and spin-2 BECs, 
there are two ($a_0$ and $a_2$) and three ($a_0$, $a_2$ and $a_4$) scattering lengths, respectively, corresponding
to total spin 0 ($a_0$), 2 ($a_2$) and 4 ($a_4$) channels \cite{review-spinor-cond}.
A spin-1 spinor BEC can be magnetically classified as either 
ferromagnetic $(a_{0}>a_{2})$ or antiferromagnetic (polar) $(a_{0}<a_{2})$. In various domains of phase
space governed by $a_0$, $a_2$ and $a_4$, a spin-2 spinor BEC manifests in three magnetic phases: ferromagnetic,
antiferromagnetic, and cyclic. These scattering lengths are close to one another in many spinor BECs
\cite{Scat-lengths-1,Scat-lengths-2}. A nonmagnetic phase of a spinor BEC corresponds to the case where these 
scattering lengths are equal. This can be achieved by engineering the scattering lengths by
{manipulating} an external electromagnetic field in the neighborhood of a Feshbach resonance(s) \cite{Feshbach}.

In this Letter{,} we study the   spatially-periodic states in quasi-2D SO-coupled nonmagnetic spin-1 
and spin-2 BECs trapped along one of the axes but free in the plane perpendicular to it.   
We will consider $a_0 = a_2 = a_4$, while, in the absence of SO coupling, a spin-1 or spin-2 BEC becomes nonmagnetic with no magnetic behavior. 
The resultant systems are SO-coupled three- (spin-1) and five-component (spin-2) BECs without any spinor
interactions. When $\gamma \lessapprox 2$, i.e. SO-coupling strength is small, the ground and the
first excited states of such a  spin-1 BEC are the spatially-periodic radially-symmetric $(-1,0,+1)$ and 
$(0,+1,+2)$ states, respectively, and the same for a spin-2 BEC are $(-2,-1,0,+1,+2)$ and $(-1,0,+1,+2,+3)$ states, respectively,
where the numbers in the parenthesis are the winding numbers in the respective spin components. For a larger 
$\gamma$, $(-1,0,+1)$ and $(-2,-1,0,+1,+2)$ states develop a multiring structure and continue as the ground states.
{Trapped spin-1 spinor BEC under rotation or azimuthal gauge potential also have similar states 
with vorticity as the ground states \cite{kita,3D-expt}}.
{
Hence, one may conclude that the complicated spinor interactions in spin-1 and spin-2
BECs are necessary for generating these states with intrinsic vorticity. In this Letter, we demonstrate that such states can 
also be generated without spinor couplings in SO-coupled BECs. In fact, the appearance of these spatially-periodic states in a much simpler  model  will allow the study of these supersolid-like states  easily  without the complications of spinor interactions and provide a better understanding of the origin of these states.}
For medium $\gamma$ ($2 \lessapprox \gamma \lessapprox 4$), two new spatially-periodic excited states $-$ 
stripe and square-lattice states $-$ appear in addition to a circularly-asymmetric state. 
The stripe state has a stripe pattern in component densities, whereas in the square-lattice 
state, total as well as component densities display square-lattice pattern, as in a supersolid. 
Additionally, in a spin-2 
BEC, for intermediate strengths of SO-coupling ($\gamma \approx 1$), a spatially-periodic supersolid-like
triangular-lattice state with a triangular 
lattice formation in component as well as total densities is also found. 
The supersolid \cite{sprsld} is a unique quantum state of matter in the sense that, 
like a superfluid, it breaks continuous gauge invariance, flows without friction, and simultaneously
breaks continuous translational invariance via the formation of stable, spatially-organized structures 
like a crystalline solid.
In other words, it defies the notion that friction-free movement is a quality reserved for quantum fluids, 
such as a BEC \cite{bose} and a Fermi superfluid \cite{fermi}, and, instead, exhibits characteristics of both 
a superfluid and a solid. The elusive supersolid state of matter has been observed in a series
of experiments with dipolar BECs which, on weakening the contact-interaction strength, evolves by 
{a} phase transition to a supersolid phase first and then {by a} crossover to an insulating phase \cite{Bottcher,Natale}. 
A supersolid-like stripe phase with coexisting phase coherence and crystalline structure which manifests as
the stripes in the total density has been seen in SO-coupled pseudospin-1/2 spinor condensates \cite{Ketterle}.
In resemblance to a supersolid, these stripes have also been referred to as superstripes in the literature \cite{Putra,Li}. 

In Sec. \ref{section2}, we introduce the coupled Gross-Pitaevskii 
(GP) equations to study the SO-coupled nonmagnetic spin-1 and spin-2 BECs in mean-field approximation 
at temperatures very close to absolute zero and later discuss the solutions in the absence of spinor interactions.
In
Sec.~\ref{section3}, we describe a variety of self-trapped solutions that result from a numerical
solution of the GP equations at various SO-coupling strengths. 
The numerical
results for a nonmagnetic spin-1 SO-coupled BEC are discussed in Sec.~\ref{section3a} whereas in 
Sec.~\ref{section3b}, we  discuss the same of an SO-coupled nonmagnetic spin-2 BEC. 
We conclude with a summary of results in Sec.~\ref{section4}.

\section{The mean-field models} 
\label{section2}
We consider a spin$-f$ condensate consisting of $N$ atoms of mass $m$ each with $a_0=a_2$ for
$f=1$ and $a_0=a_2=a_4$ for $f=2$. The condensate is trapped axially by a harmonic potential $V({\bf r})= m\omega_z^2 z^2/2$, where $\omega_z$ is the angular trap frequency.  The single-particle Hamiltonian 
of the  quasi-2D Rashba SO-coupled BEC  is obtained by integrating out the $z$ coordinate \cite{thspinorb,exptso} as
\begin{equation}
\label{SPH}
H_{0} = -\mathbb{1}\frac{\hbar ^{2}}{2 m}\nabla _{\boldsymbol{\rho }}^{2}+ 
\gamma(S_x\hat{p}_y-S_y\hat{p}_x),\\
\end{equation}
where ${\boldsymbol{\rho }}\equiv \{x,y \}$,
$\nabla _{\boldsymbol{\rho }}^{2}= (\partial _{x}^{2}+\partial _{y}^{2})$ with
$\partial _{x} = \partial /\partial x$ and
$\partial _{y} = \partial /\partial y$, 
$\hat{p}_x =-i\hbar \partial _{x}$ and $\hat{p}_y=-i\hbar \partial _{y}$ are, respectively, 
$x$ and $y$ components of the momentum operator, $\gamma$ is the strength of Rashba SO coupling, $\mathbb{1}$
represents a $(2f+1)\times (2f+1)$ identity matrix, and $S_{x}$ and $S_{y}$ are, respectively, the 
$x$ and $y$ components of angular momentum operator ${\bf S}$ in irreducible representation for a spin-$f$ system. The $(j',j)^{th}$ element
of these $(2f+1)\times(2f+1)$ matrices are
\begin{subequations}
\begin{align}
(S_x)_{j',j} =& \frac{1}{2}\sqrt{(f(f+1)-j'j)}\left(\delta_{j',j+1}
                 +\delta_{j'+1,j}\right),\label{s_x}\\
(S_y)_{j',j} =&  \frac{1}{2i}\sqrt{(f(f+1)-j'j)}\left(\delta_{j',j+1}
            -\delta_{j'+1,j}\right),\label{s_y}
\end{align}
\end{subequations}
where $j'$ and $j$ can have values from $f$,$f-1,$ \ldots$, -f$. The reduced quasi-2D coupled GP equations \cite{sala} of the
SO-coupled three-component spin-1 BEC are \cite{thspinorb}
\begin{subequations}
\begin{align}
i \partial_t&\psi_0({\boldsymbol \rho})= {\cal H} \psi_{0}({\boldsymbol \rho})
{-i\frac{\gamma}{\sqrt{2}} [\partial_{-} \psi_{+1}  ({\boldsymbol \rho})  +  \partial_{+} \psi_{-1} ({\boldsymbol \rho}) ]},\label{EQ1}\\
i \partial_t &\psi_{\pm 1}({\boldsymbol 
\rho})= {\cal H}   \psi_{\pm 1}({\boldsymbol 
\rho})
-i \frac{\gamma}{\sqrt{2}}\partial_{\pm}  \psi_{0}  ({\boldsymbol \rho}),\label{EQ2} 
\end{align}
\end{subequations}
and those of the SO-coupled five-component spin-2 BEC are \cite{thspinorb}
\begin{subequations}
\begin{align}
i\partial_t\psi_0{(\boldsymbol 
\rho)} =& \mathcal{H} \psi_0 (\boldsymbol\rho)-i\textstyle\sqrt{\frac{3}{2}}\gamma[{\partial_{-}\psi_{+1}(\boldsymbol 
\rho)} + {\partial_{+}\psi_{-1}(\boldsymbol \rho)} ], \label{cgpet3d-1}\\
i\partial_t\psi_{\pm 1}{(\boldsymbol \rho)} =& \mathcal{H} \psi_{\pm 1}(\boldsymbol \rho)
-i\gamma\left[ \textstyle \sqrt{\frac{3}{2}}{\partial_{\pm}} \psi_0(\boldsymbol \rho) 
+{\partial_{\mp}} \psi_{\pm 2}(\boldsymbol \rho)\right] , 
\label{cgpet3d-2}\\
i\partial_t\psi_{\pm 2}{(\boldsymbol \rho)} =& \mathcal{H} \psi_{\pm 2}(\boldsymbol \rho)- 
i\gamma\partial_{\pm}\psi_{\pm 1}(\boldsymbol \rho). 
\label{cgpet3d-3}
\end{align}
\end{subequations}
In Eqs.~(\ref{EQ1})-(\ref{EQ2}) and Eqs. (\ref{cgpet3d-1})-(\ref{cgpet3d-3}), 
${\cal H}=-\nabla^2_{\boldsymbol \rho}/2+ c_0 n$, $c_0 = 2{N}\sqrt{2\pi}a_0,$
$\partial _{t} = \partial /\partial t$, $\partial_{\pm} = \partial_{y} \pm i \partial_{x}$, 
$n ({\boldsymbol{\rho }})= \sum _{j} n_{j}({\boldsymbol{\rho }})$ is the total density, where 
$n_{j} = |\psi _{j}|^{2}$ are the densities of components $j=\pm 1,0$ for $f=1$ and $j=\pm 2,\pm 1,0$
for $f=2$. In Eqs.~{(\ref{EQ1})}-{(\ref{EQ2}) and} Eqs.~{(\ref{cgpet3d-1})}-{(\ref{cgpet3d-3})}, 
length is expressed in units of  $l_{0}= \sqrt{\hbar / m\omega _{z}}$, wave function in units of  $l_{0}^{-1}$, $t$ in units of $\omega _{z}^{-1}$ and energy in units of $\hbar \omega _{z}$;
the order parameter satisfies the normalization condition 
{
\begin{align}\label{norm}
{\textstyle \int }n({\boldsymbol{\rho }})\, d{\boldsymbol{\rho }}\equiv {\textstyle \int \sum_j }|\psi_j({\boldsymbol{\rho }})|^2\, d{\boldsymbol{\rho }}
=1.
\end{align} 
{Although, spin-exchange spinor interaction is absent in nonmagnetic spin-1 and spin-2 condensates, the SO-coupling interaction will allow a mixing of different components and the  number of atoms in each component will not be conserved in this case.}}
The  energy functionals corresponding to the  
mean-field GP equations (\ref{EQ1})-(\ref{EQ2}) for a spin-1 and   (\ref{cgpet3d-1})-(\ref{cgpet3d-3}) for a spin-2 BEC 
are, respectively,
\begin{align}
\label{energy}
E[\psi] &= {\textstyle \frac{1}{2} \int }d{\boldsymbol{\rho }} \big [{\textstyle \sum}_{j=-1}^{+1} 
|\nabla _{\boldsymbol{\rho }}\psi _{j}|^{2}+c_{0}n^2-\sqrt{2}i\gamma\big \{ \psi _{+1}^{*}
 \partial_{+}\psi_{0}\nonumber \\
&
 +\psi_{0}^{*} \partial_{-}\psi_{+1}+ \psi_{0}^{*}
\partial_{+}\psi_{-1} + \psi_{-1}^{*}\partial_{-}\psi_{0}\big\} \big ],\qquad
\end{align}
and 
\begin{align}
\label{energy-spin2}
E[\psi] &= {\textstyle \frac{1}{2}  \int d{\boldsymbol{\rho }} \Big[ {
\textstyle \sum }_{j=-2}^{+2} |\nabla_{\boldsymbol{\rho}}\psi_{j}|^{2}+c_{0}n^{2}-2i\gamma\Big\{\psi_{+2}^{*}
\partial_{+}\psi_{+1} }  \nonumber \\
&+\psi_{+1}^{*}\Big(\textstyle \sqrt{\frac{3}{2}}{\partial_{+}}\psi_0 + {\partial_{-}} 
\psi_{+2}\Big)+\psi_{0}^{*}\textstyle\sqrt{\frac{3}{2}}\Big({\partial_{-}\psi_{+1}} + {\partial_{+}\psi_{-1}}\Big)
\nonumber \\
&+\psi_{-1}^{*}\Big(\textstyle \sqrt{\frac{3}{2}}{\partial_{-}}\psi_0 + {\partial_{+}} 
\psi_{-2}\Big)+\psi_{-2}^{*}\partial_{-}\psi_{-1}\Big\}\Big].
\end{align}

 In the absence of interactions, Eqs.~{(\ref{EQ1})}-{(\ref{EQ2})} for $f = 1$ or Eqs.~{(\ref{cgpet3d-1})}-{(\ref{cgpet3d-3})} for $f=2$ reduce to an eigen value problem for the 
single-particle Hamiltonian $H_0$ in Eq.~(\ref{SPH}) which is exactly solvable. The minimum energy eigen
function and eigen energy of $H_0$ for $f = 1$ are
\begin{align}
\Phi_{f=1} &= \frac{1}{2}\begin{pmatrix}
-e^{-i\varphi} \\
-\sqrt{2}i\\
 e^{i\varphi}\\
\end{pmatrix}e^{ixk_x+iyk_y} \equiv \zeta(\varphi)e^{ixk_x+iyk_y}, 
\label{ef_sph}\\
E(k_x,k_y) &= \frac{1}{2}\left(k_x^2+k_y^2-2\gamma\sqrt{k_x^2+k_y^2}\right),\label{minima}
\end{align}
respectively, where $\varphi$ = $\tan^{-1}(k_{y}/k_{x})$ denotes the orientation of propagation vector
${\bf k} = (k_x,k_y)$. The magnitude of the propagation vector is fixed by minimizing the dispersion
in  Eq. (\ref{minima}) which gives $k_{x}^2+k_{y}^2 = \gamma^2$.
All the plane-wave eigenfunctions in Eq. ({\ref{ef_sph}}) with different $\varphi$ are degenerate.
The equal-weight superposition of these degenerate plane-wave states yields
\begin{align}\label{abcd}
\Phi^{f=1}_{\rm MR} &= \frac{1}{4\pi}\int_{0}^{2 \pi}\begin{pmatrix}
-e^{-i\varphi} \\
-\sqrt{2}i\\
 e^{i\varphi}\\
\end{pmatrix}e^{i\gamma r\cos(\varphi-\theta)}d\varphi,\\
&= \frac{1}{2}\begin{pmatrix}
- i e^{- i \theta } J_{+1}( \gamma r)\\
-\sqrt{2} i J_0( \gamma r)\\
 ie^{i \theta } J_{-1}( \gamma r) \\
 \end{pmatrix},
\label{gen_ef_sph1}
\end{align}
where polar coordinate $\theta = \tan^{-1}{y/x}$, $J_{0}(\gamma r)$ and $J_{\pm 1}(\gamma r)$ are the 
Bessel functions of first kind of order $0$ and $\pm 1$, respectively. The function
$\Phi_{\rm MR}^{f=1}$ carries the phase singularities with the same winding numbers as a $(-1,0,+1)$-type multiring (MR) soliton.
Rather than considering the superposition of infinite plane waves as in Eq.~(\ref{gen_ef_sph1}),
the superposition{s} of a finite number of  plane waves like a pair of
counter-propagating plane waves, or of
three plane waves propagating at mutual angles of $2\pi/3$, or of
 two pairs of counter-propagating plane waves with propagation vectors of one pair perpendicular to the other are 
\begin{subequations}\label{phi_sl}
 \begin{align}
  \Phi^{f=1}_{\rm ST} &= \textstyle \frac{1}{\sqrt{2}}\left[\zeta(0)e^{i\gamma x}
                 + \zeta(\pi)e^{-i\gamma x}\right],\\
  \Phi^{f=1}_{\rm TL} &=\textstyle \frac{1}{\sqrt{3}} \left[\zeta(0)e^{i\gamma x}
                 + \zeta(2\pi/3)e^{i\gamma(-x+\sqrt{3}y)/2}\right.
                   \nonumber\\
                &\left.+\zeta(4\pi/3)e^{i\gamma(- x- \sqrt{3} y)/2}\right],\\
  \Phi^{f=1}_{\rm SL} &= \textstyle \frac{1}{2}\left[\zeta(0)e^{i\gamma x}
                +\zeta(\pi/2)e^{i\gamma y}+\zeta(\pi)e^{-i\gamma x}\right.
                \nonumber\\
                &\left.+\zeta(3\pi/2)e^{-i\gamma y}\right],
 \end{align}
\end{subequations}
respectively. The total density corresponding $\Phi^{f=1}_{\rm ST}$, $\Phi^{f=1}_{\rm TL}$, and 
$\Phi^{f=1}_{\rm SL}$ have stripe (ST), triangular lattice (TL), and square lattice (SL) patterns, 
respectively.

Similarly for a spin-2 BEC, the eigenfunction corresponding to the minima of the lowest dispersion branch 
at $k_{x}^2+k_{y}^2 = (2\gamma)^2$ and with eigen energy  $-2\gamma^2$ is \cite{SS-spin-2}
\begin{equation}
\Phi^{f=2} = \frac{1}{4}\begin{pmatrix}
e^{-2i\varphi} \\
2 i e^{-i\varphi}\\
-\sqrt{6} \\
-2 i e^{i\varphi}\\
e^{2i\varphi}\\
\end{pmatrix}e^{ixk_x+iyk_y} \equiv \eta(\varphi)e^{ixk_x+iyk_y}.
                                  \label{ef_sph-2}
\end{equation}
The solution corresponding to superposition of infinite pairs of counter-propagating
plane waves is \cite{SS-spin-2} 
\begin{align}
\Phi^{f=2}_{\rm MR} 
&= \frac{1}{4}\begin{pmatrix}
-e^{-2 i \theta } J_{+2}( 2 \gamma r)\\
-2 e^{- i \theta } J_{+1}( 2 \gamma r)\\
-\sqrt{6}J_0( 2 \gamma r)\\
 2 e^{ i \theta } J_{-1}(2 \gamma r)\\
-e^{2 i \theta } J_{-2}( 2 \gamma r)\label{gen_ef_sph}
\end{pmatrix},
\end{align}
and the superpositions which yield a stripe (ST), triangular lattice (TL),
and square lattice (SL) density patterns, respectively, are \cite{SS-spin-2}
\begin{subequations}\label{phi_sl-2}
 \begin{align}
  \Phi^{f=2}_{\rm ST} &= \textstyle \frac{1}{\sqrt{2}}\left[\eta(0)e^{i2\gamma x}
                 + \eta(\pi)e^{-i2\gamma x}\right],\\
  \Phi^{f=2}_{\rm TL} &=\textstyle \frac{1}{\sqrt{3}} \left[\eta(0)e^{i2\gamma x}
                 + \eta(2\pi/3)e^{i\gamma(-x+\sqrt{3}y)}\right.
                   \nonumber\\
                &\left.+\eta(4\pi/3)e^{i\gamma(-x-\sqrt{3}y)}\right],\\
  \Phi^{f=2}_{\rm SL} &= \textstyle \frac{1}{2}\left[\eta(0)e^{i2\gamma x}
                +\eta(\pi/2)e^{i2\gamma y}+\eta(\pi)e^{-i2\gamma x}\right.
                \nonumber\\
                &\left.+\eta(3\pi/2)e^{-i2\gamma y}\right].
 \end{align}
\end{subequations}
In the following, we will see in our numerical study that these three types of states $-$ stripe, triangular-lattice
and square-lattice $-$ indeed appear in the spin-2 case. However in the spin-1 case we could only identify stripe and
square-lattice states.

\section{Numerical Results}
\label{section3}
We numerically solve Eqs.~{(\ref{EQ1})}-{(\ref{EQ2})} for $f = 1$ and Eqs.~(\ref{cgpet3d-1})-(\ref{cgpet3d-3}) 
for $f=2$ over a finite spatial domain, given an initial solution and boundary conditions, using Crank-Nicolson
\cite{bec2009}
or Fourier pseudo-spectral methods \cite{Spin1and2our} complimented with time-splitting to treat the non-linear terms more 
efficiently. To calculate stationary-state solutions,  Eqs.~{(\ref{EQ1})}-{(\ref{EQ2})} and Eqs.~(\ref{cgpet3d-1})-(\ref{cgpet3d-3}) are first transformed
by changing $t \rightarrow -i \bar{t}$ and then solved by imaginary-time propagation \cite{Spin1and2our,bec2009a} using appropriate initial guesses. In order
to confirm the dynamical stability of a solution, we study its real-time propagation \cite{Spin1and2our,bec2009a}  as governed by 
the coupled GP equations. The space step used in the calculation was typically $\Delta\sim 0.05$ and the time step was $\sim 0.1\Delta^2$ in imaginary-time propagation and $\sim 0.05\Delta^2$ in real-time propagation.  {The SO coupling will allow mixing between different spin components and, hence, the number of atoms in different components will not be conserved during 
{imaginary-time} propagation in numerical calculations.  Nevertheless, the total number of atoms will be preserved and we impose condition (\ref{norm}) during time propagation. As the number of atoms in each component is not conserved, the final conserved result is independent of the initial choice of number of atoms in each component.  }

\subsection{Nonmagnetic quasi-2D solitons in an SO-coupled spin-1 BEC}

\label{section3a}

\begin{figure}[!t] 
\centering
\includegraphics[width=\linewidth]{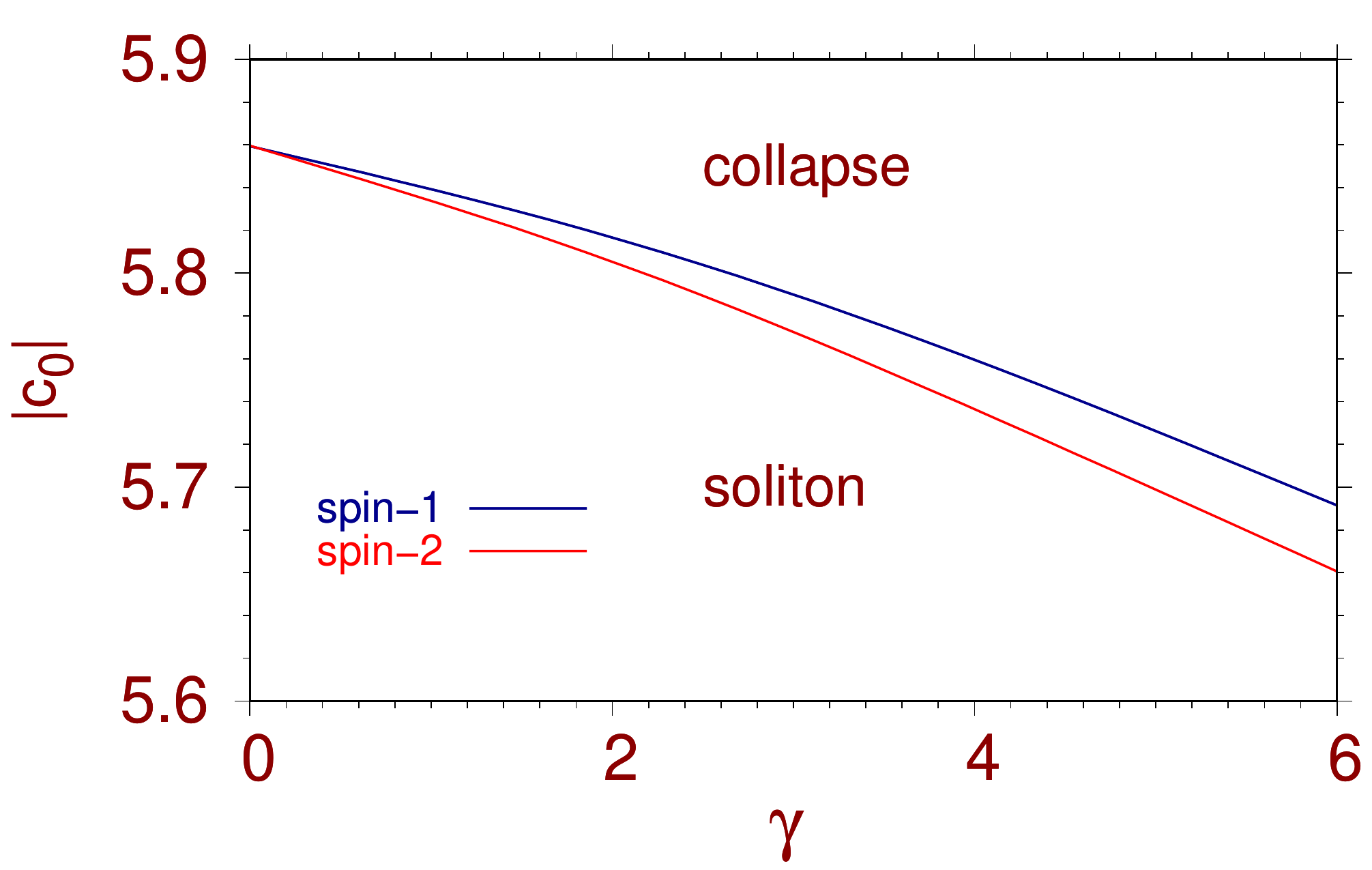}
\caption{The $c_0$-$\gamma$ phase plot illustrating the collapse and  soliton formation. }
\label{fig1a}

\end{figure}

\begin{figure}[!t] 
\centering
\includegraphics[width=\linewidth]{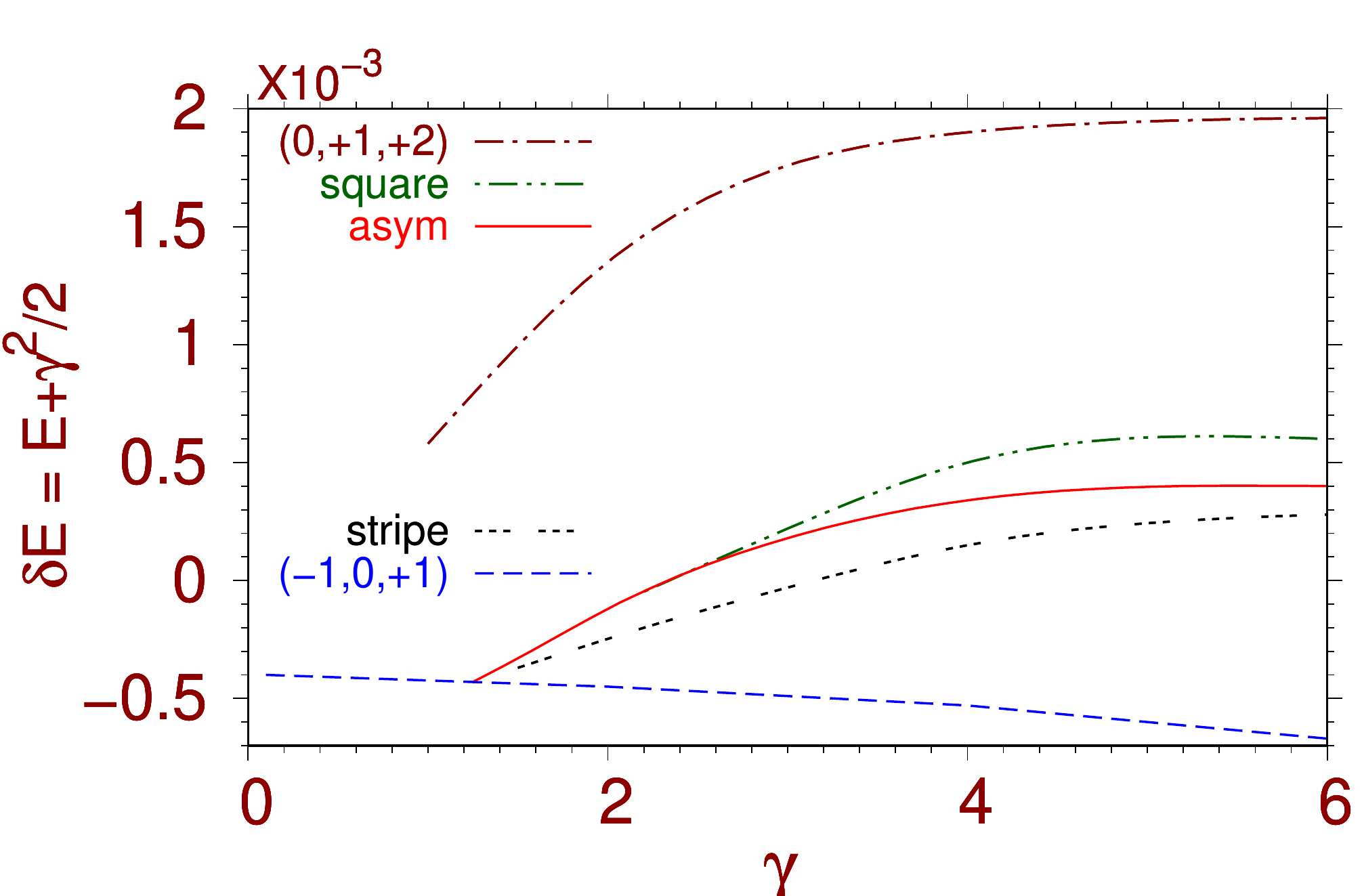}
\caption{ The $\delta E \equiv E+\gamma^2/ 2$ versus $\gamma$  phase plot for Rashba  SO coupled 
  $(-1,0,+1)$-, $(0,+1,+2)$-type multiring, asymmetric, stripe and square-lattice solitons.}
\label{fig1}

\end{figure}

To obtain quasi-2D solitons with a self-attractive ($c_{0}<0$) nonmagnetic SO-coupled spin-1 BEC for different 
SO-coupling strengths $\gamma$, we consider a BEC with $c_{0}=-0.6$ and vary $\gamma$ in this study. We
consider $c_{0}=-0.6$ because this value gives an adequate size of the soliton. A decrease in $c_0$ results 
in more attraction, hence a reduced system size,  { and below a critical $c_0=c_{0,\mathrm{crit}}$ the BEC will collapse as displayed in Fig. \ref{fig1a}.
For $\gamma=0$, the spin-1 and spin-2 systems become essentially identical and also equivalent to a scalar nonspinor BEC of $N$ atoms. The nonlinearities in the mean-field equations in these three cases are also identical. 
For $\gamma=0$, for a nonspinor spin-0 BEC, the critial $c_0$ for collapse was obtained previously as $c_{0,\mathrm{crit}}\sim -5.85$ \cite{malomed,dali}.  
 In this study we find, for spin-0, spin-1 and spin-2 cases   and  for $\gamma=0$,  $c_{0,\mathrm{crit}}= -5.86$, quite close to this  previous   spin-0 result. The $c_{0,\mathrm{crit}}$ for collapse  for nonzero $\gamma$ for spin-1 and spin-2 cases are slightly different as illustrated in Fig. \ref{fig1a}.
On the other
hand, an increase in $c_0$ leads to an increased system size and, for positive (self-repulsive) values of $c_0$, the system is 
no longer self-trapped. For negative (self-attractive) values of $c_0$ above the critical value ($c_0>c_{0,\mathrm{crit}}$), the SO-coupled BEC  remains self-trapped.}
 We demonstrate in Fig.~\ref{fig1} how different types of solitons appear as SO-coupling strength is varied for a fixed $c_0$ through a phase plot of energy $\delta E \equiv (E+\gamma^2/2)$ against $\gamma $, where $E$ denotes the energy of the soliton and $-\gamma^2/2$ corresponds to the energy of single particle Hamiltonian (\ref{SPH}) for a spin-1 BEC.
We find five different types of quasi-2D
solitons for different $\gamma$: (a) $(-1,0,+1)$ and $(0,+1,+2)$ multiring solitons, (b) stripe soliton with a stripe
pattern in component densities only, (c) 
a supersolid-like soliton with component and total densities having 
the square-lattice pattern,
and (d) asymmetric soliton. From Fig. \ref{fig1}, we find that for small $\gamma$
($\gamma \lessapprox 1 \sim 2$) only $(-1,0,+1)$- and $(0,+1,+2)$-type multiring solitons are possible 
with the $(0,+1,+2)$ soliton being an excited state. For medium $\gamma$ ($3\gtrapprox \gamma \gtrapprox 1.5)$, two 
new types of excited states appear: stripe and asymmetric solitons; of these the stripe soliton has a smaller energy 
than the asymmetric soliton. For large $\gamma$ ($\gamma \gtrapprox 3$) the square-lattice excited state appears 
with an energy larger than the asymmetric soliton. 

\begin{figure}[!t] 
\centering
\includegraphics[width=\linewidth]{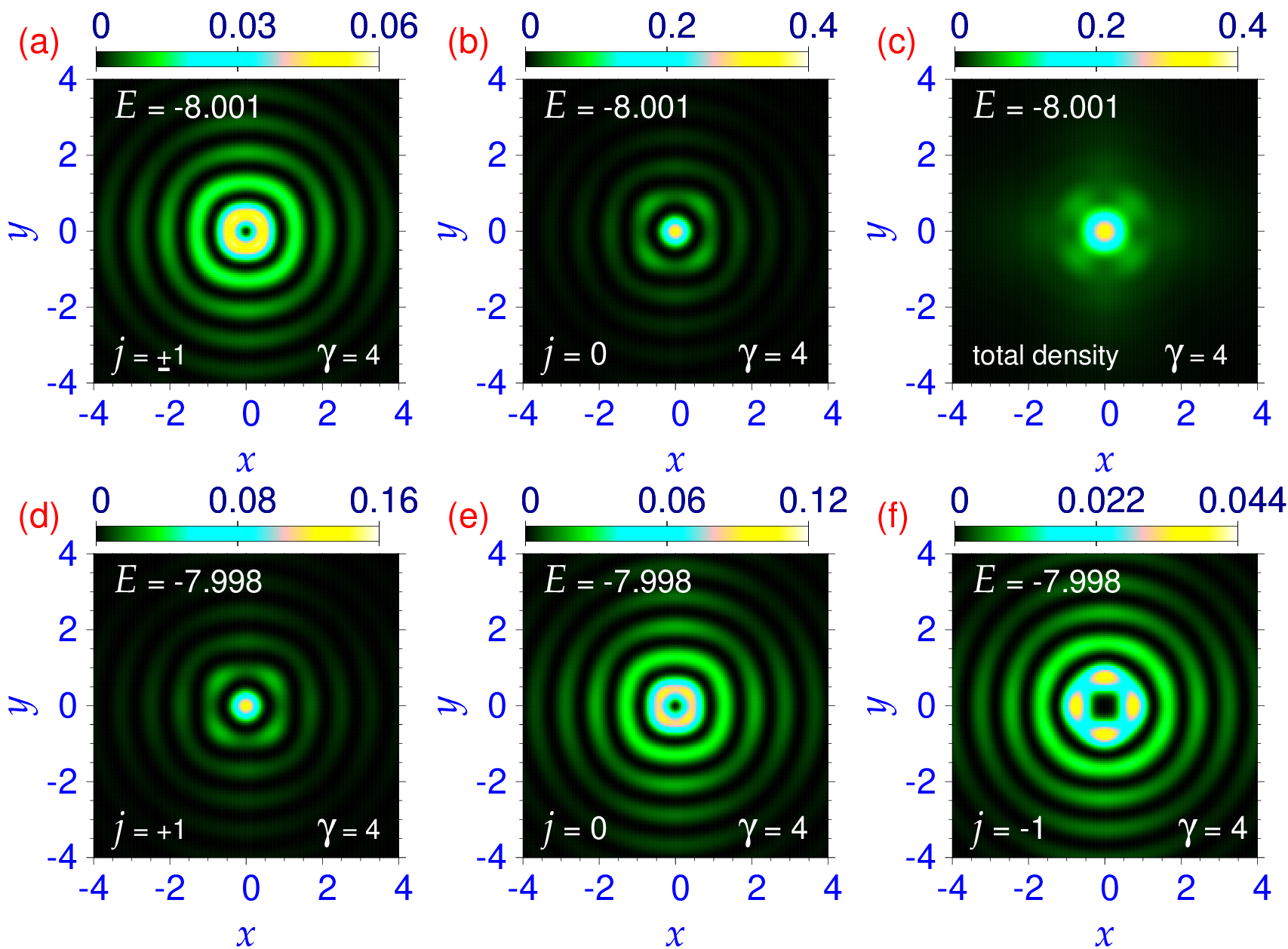}
\includegraphics[width=\linewidth]{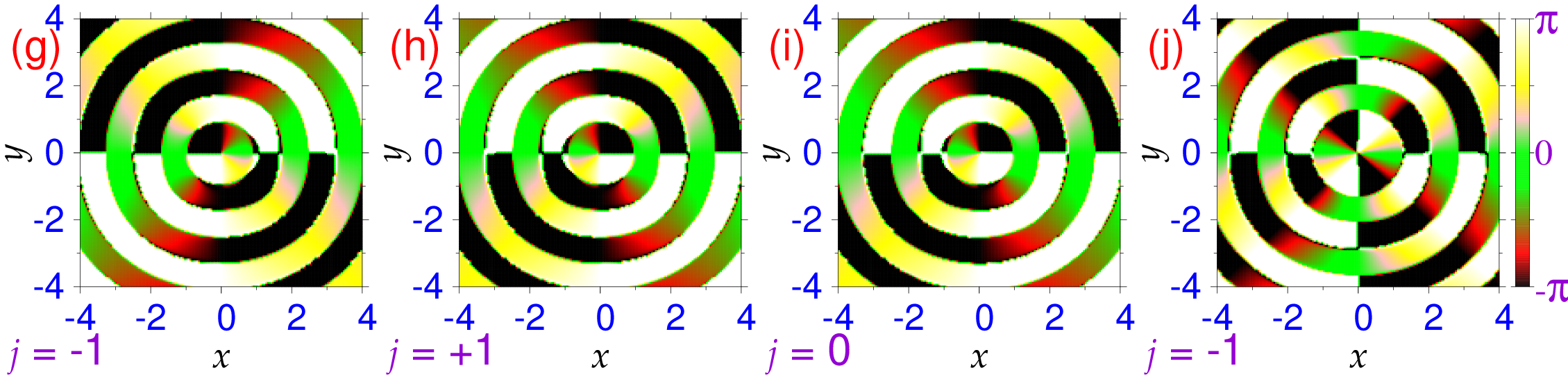}
\caption{Top row: (a) component density $n_{\pm 1}(\boldsymbol \rho)$, (b)  $n_0(\boldsymbol \rho)$,
and (c) total density  $n(\boldsymbol \rho)$ {of} a $(-1,0,+1)$-type multiring soliton of the SO-coupled spin-1
BEC with $c_0=-0.6$ and $\gamma=4$. Similarly, middle row: (d) component density $n_{+1}(\boldsymbol \rho)$,
(e) $n_0(\boldsymbol \rho)$, and (f) $n_{-1}(\boldsymbol \rho)$ {of} a $(0,+1,+2)$-type multiring soliton of
the BEC with same interaction and coupling strengths. Bottom row: phase distributions of wave-function components 
(g) $\psi_{+1}$, (h)  $\psi_{-1}$ {of} the $(-1,0,+1)$-type multiring BEC soliton and  those of wave-function components (i) $\psi_0$,
(j) $\psi_{-1}$ {of} the  $(0,+1,+2)$-type multiring BEC soliton.
}
\label{fig2}
\end{figure}

\begin{figure}[!t] 
\centering
\includegraphics[width=\linewidth]{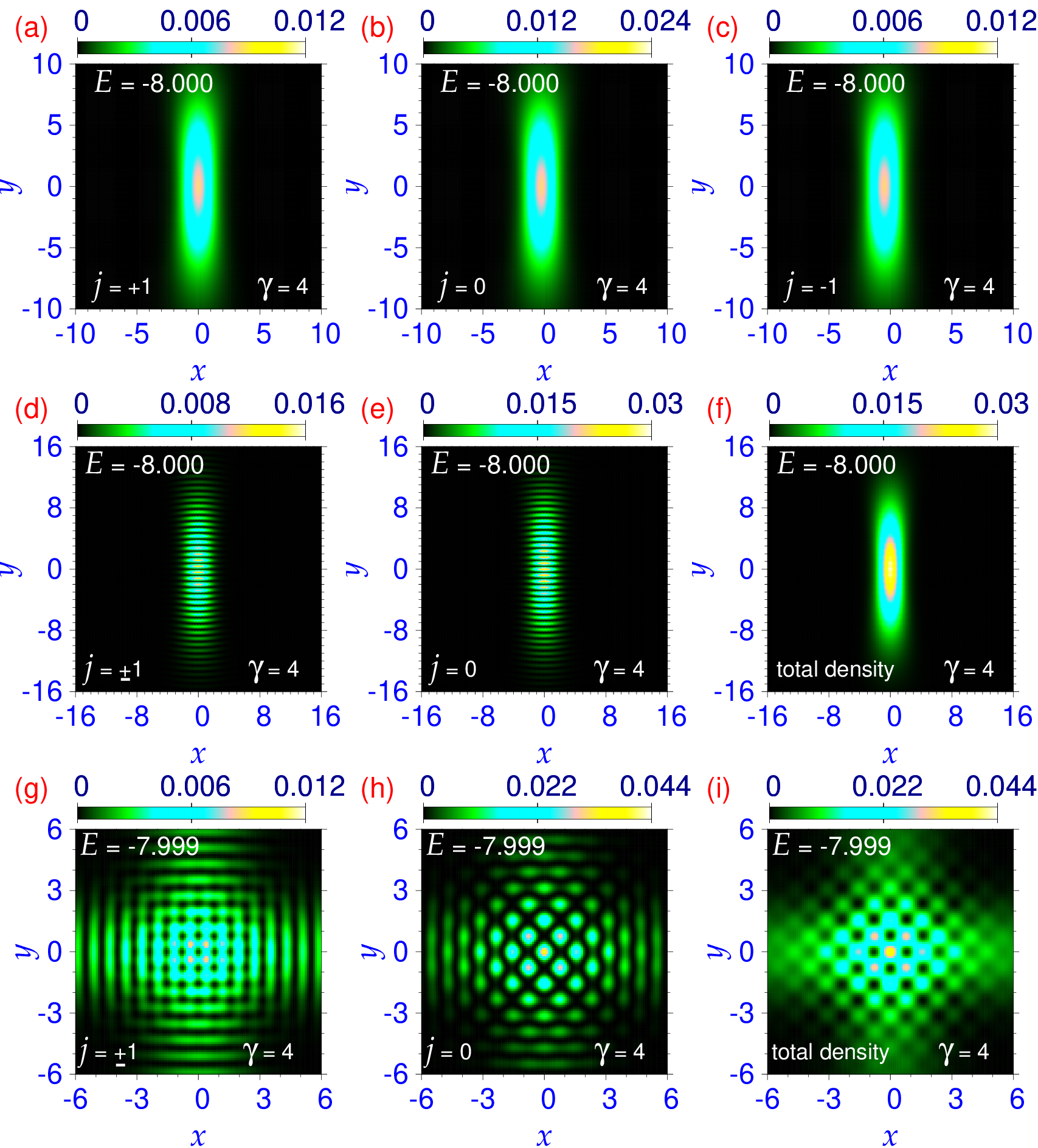} 
\caption{Top row: (a) component density $n_{+ 1}(\boldsymbol \rho)$, (b)  $n_0(\boldsymbol \rho)$,
and (c) $n_{-1}(\boldsymbol \rho)$ {of} an asymmetric soliton of the SO-coupled spin-1 BEC with 
$c_0=-0.6$ and $\gamma=4$. Middle row: (d) component density $n_{\pm 1}(\boldsymbol \rho)$, (e)  
$n_0(\boldsymbol \rho)$, and (f) total density  $n(\boldsymbol \rho)$ {of} a stripe soliton and bottom row:
(g) component density $n_{\pm 1}(\boldsymbol \rho)$, (h)  $n_0(\boldsymbol \rho)$, and (i) total density  $n(\boldsymbol \rho)$ 
{of} a square-lattice soliton for the same interaction and coupling strengths. }
\label{fig3}
\end{figure}

For the Rashba SO coupling, we first investigate the formation of two quasi-degenerate multiring 
vector solitons of $(- 1, 0,+ 1)$- and $(0,+ 1,+ 2)$-type generated by solving Eqs.~(\ref{EQ1})-(\ref{EQ2}) in 
imaginary-time using initial guesses where appropriate vortices are imprinted in the components. We show the 
contour plots of the densities (a) $n_{\pm 1}(\boldsymbol \rho)$, (b) $n_{0}(\boldsymbol \rho)$, and (c) 
$n(\boldsymbol \rho)$ (total density) in a $(- 1,0,+ 1)$-type multiring soliton of the spin-1 BEC with 
$c_{0}=-0.6$ and  $\gamma = 4$ in Figs.~\ref{fig2}(a)-(c). 
For the same parameters, we show the densities of three components, $n_j(\boldsymbol \rho)$, in a $(0, +1, +2)$-type multiring 
soliton in Figs.~\ref{fig2}(d)-(f). The energies of these $(- 1,0,+ 1)$- and $(0,+ 1, + 2)$-type solitons are 
$E=-8.001$ and $E=-7.999,$ respectively; the latter is an excited state. The phase distributions 
of $\psi_{+1}$ and $\psi_{-1}$ of  the $(-1,0,+1)$-type multiring soliton shown in Figs.~\ref{fig2}(g) and (h) 
reflect the phase shifts of $-2\pi$ and $2\pi$, respectively, under a full rotation around the center.
Similarly, phase shifts by $2\pi$ and $4\pi$, respectively, under a full rotation around the center of $\psi_0$ and $\psi_{-1}$ of the $(0,+1,+2)$-type multiring soliton
 can be seen in  Figs.~\ref{fig2}(i) and (j). These phases agree with the vortex/anti-vortex structure of the $(-1,0,+ 1)$ and $(0,+ 1,+ 2)$-type solitons. 

Besides the two circularly-symmetric solitons, asymmetric, stripe and square-lattice
solitons are other stationary states of the BEC with $c_{0} =-0.6$ and $\gamma =4$. The component densities
$n_j(\boldsymbol \rho )$ of the asymmetric, stripe and square-lattice  solitons are shown, respectively, 
in Figs.~\ref{fig3}(a)-(c), \ref{fig3}(d)-(f), and \ref{fig3}(g)-(i). The energies of these three are $-8.000$,
$-8.000$ and $-7.999$, respectively. The stripe soliton has a stripe pattern in the density 
$n_j(\boldsymbol \rho)$ of each component, whereas the total density $n(\boldsymbol \rho)$ is devoid of this pattern. The square-lattice soliton, on other hand, has a square 
lattice pattern in each component density $n_j(\boldsymbol \rho)$ and also in the total density 
$n(\boldsymbol \rho)$. The energies of the different solitons 
for $\gamma =4$ satisfy $E(0,+1,+2) > E$(squarelattice) $>E$(asymmetric) $>E$(stripe) $>E(-1,0,+1)$. The stripe 
and the square-lattice solitons are efficiently obtained in numerical calculation, if such periodic patterns 
are imprinted on the initial state \cite{SS-SPIN1}

\subsection{{N}onmagnetic quasi-2D solitons in an SO-coupled spin-2 BEC} \label{section3b}
\begin{figure}[!t] 
\centering
\includegraphics[width=\linewidth]{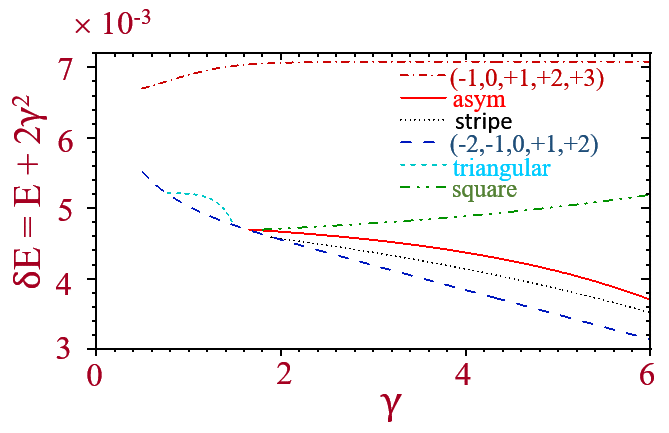}
\caption{ The $\delta E \equiv E+2\gamma^2$ versus $\gamma$  phase plot for Rashba SO-coupling 
of  $(-2,-1,0,+1,+2)$- and  $(-1,0,+1,+2,+3)$-type multiring, asymmetric, triangular-lattice, stripe and square-lattice solitons.}
\label{spin2-bi}
\end{figure}

\begin{figure}[!t] 
\centering
\includegraphics[width=1\linewidth]{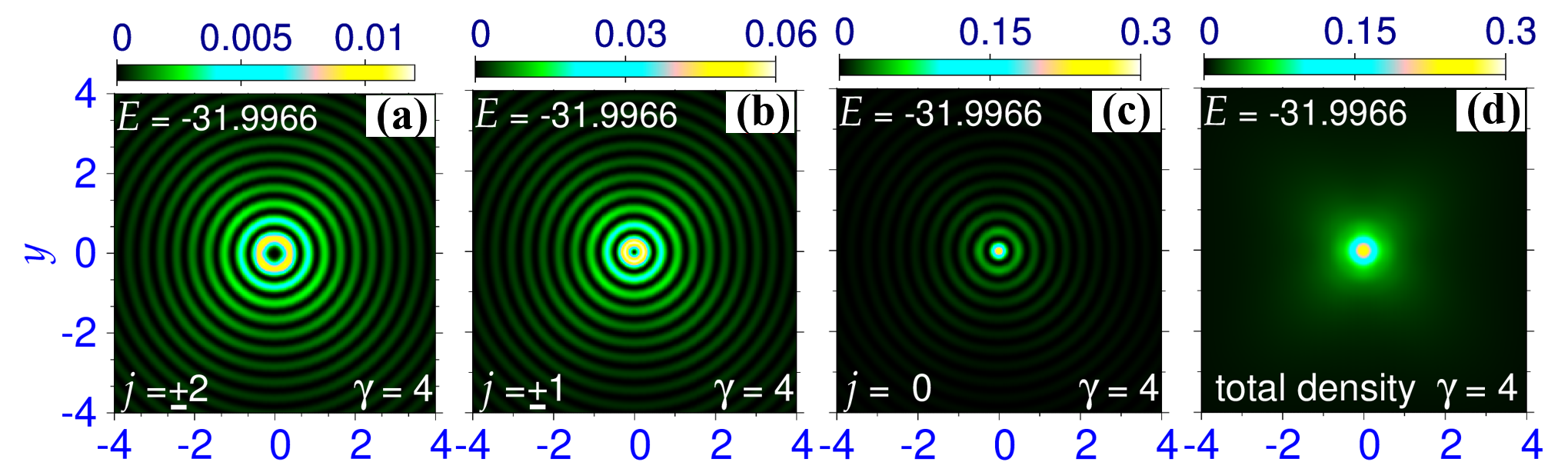} 
\hbox{\hspace{2mm}\includegraphics[width=1\linewidth]{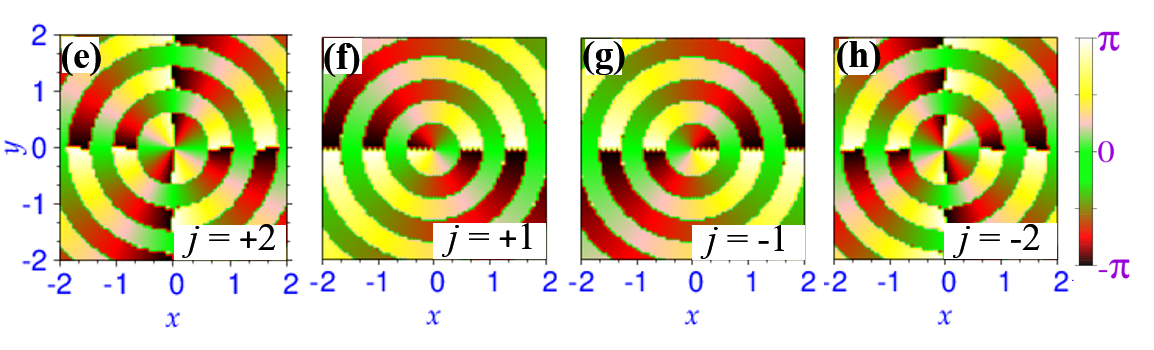}} 
\caption{Upper row: (a) component density $n_{\pm 2}(\boldsymbol \rho)$, (b)  $n_{\pm 1}(\boldsymbol \rho)$,
(c) $n_{0}(\boldsymbol \rho)$, and (d) total density $n(\boldsymbol \rho)$ {of} a $(-2,-1,0,+1,+2)$-type
multiring soliton of the SO-coupled spin-2 BEC with $c_0=-0.2$ and $\gamma=4$. Lower row:
phase distributions of wave-function components (e) $\psi_{+2}$, (f) $\psi_{+1}$, (g) $\psi_{-1}$, and
(h) $\psi_{-2}$ in the same soliton.}
\label{Figure5}
\end{figure}

\begin{figure}[!t] 
\centering
\includegraphics[width=1\linewidth]{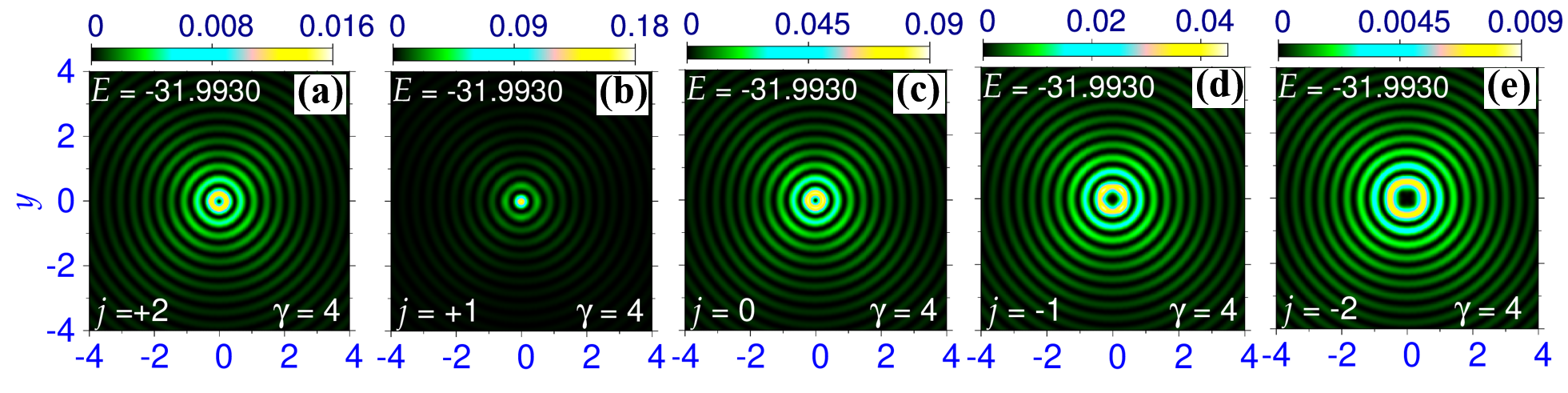} 
\includegraphics[width=1\linewidth]{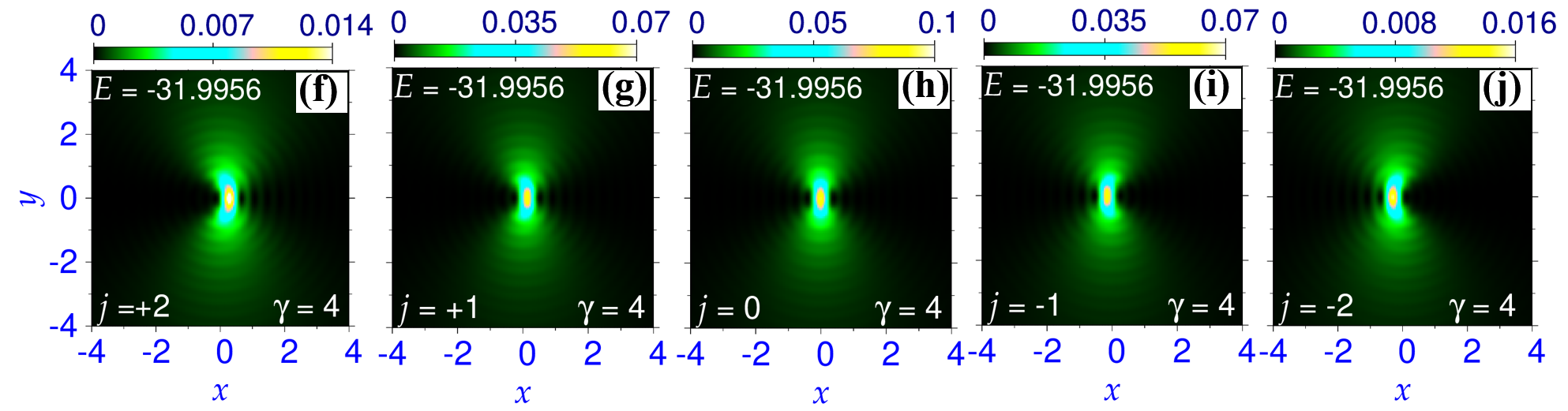} 
\caption{Upper row: (a) component density $n_{+ 2}(\boldsymbol \rho)$, (b)  $n_{+ 1}(\boldsymbol \rho)$,
(c) $n_0(\boldsymbol \rho)$, (d) $n_{-1}(\boldsymbol \rho)$, and (e) $n_{-2}(\boldsymbol \rho)$ 
{of} a $(-1,0,+1,+2,+3)$-type multiring soliton of the SO-coupled spin-1 BEC with 
$c_0=-0.2$ and $\gamma=4$. Bottom row: (f) component density $n_{+ 2}(\boldsymbol \rho)$, 
(g) $n_{+ 1}(\boldsymbol \rho)$, (h) $n_0(\boldsymbol \rho)$, (i) $n_{-1}(\boldsymbol \rho)$, and 
(j) $n_{-2}(\boldsymbol \rho)$ {of} an asymmetric soliton for the same interaction and coupling strengths.
}
\label{Figure2&3}
\end{figure}

Similar to a nonmagnetic SO-coupled spin-1 BEC, for a nonmagnetic SO-coupled spin-2 BEC too, 
 several self-trapped stationary states are possible, including those in which 
the density $n(\boldsymbol \rho)$ of the condensate displays spatially-periodic patterns. 
As in the spin-1 BEC, there are two types of
multiring solitons as well as  spatially-periodic stripe 
and square-lattice solitons.
In addition, we observe a triangular-lattice soliton that was absent in the spin-1 BEC.
For $f=2$, we consider $c_0 = -0.2$. Again, for large $|c_0|$ the system collapses as shown in Fig. \ref{fig1a}.
Here
the phase plot of $\delta E = (E+2\gamma^2$) versus $\gamma$ in Fig.~\ref{spin2-bi} shows 
how different types of solitons emerge for Rashba SO coupling, where $E$ again is the energy of the soliton 
and $-2\gamma^2$ is that of the single-particle Hamiltonian (\ref{SPH}) for a spin-2 BEC. 
We find six different types of quasi-2D solitons for different $\gamma$: (a) $(-2, −1, 0, + 1,+ 2)$- and $(- 1, 0, + 1, + 2, + 3)$-type
multiring solitons, (b) stripe soliton with stripe patterns in component densities only, (c) square-lattice soliton
having square-lattice patterns in total and all component densities{,}
(d) asymmetric soliton{,} and (e)
triangular-lattice soliton  where a triangular-lattice pattern is present in the component and total densities, similar to a supersolid. According to Fig.~\ref{spin2-bi}, multiring solitons of the types $(-2,−1, 0, +1,+2)$ and $(-1, 0, +1, +2, +3)$ are possible for small {$\gamma$ ($\gamma 
\lessapprox 1$)}.
For intermediate strengths of $\gamma$ ($\gamma \approx 1$), triangular-lattice solitons also appear as a 
quasi-degenerate state. For larger $\gamma$ ($\gamma \gtrapprox 2$), asymmetric, stripe and 
square-lattice solitons appear. These excited states in the spin-2 BEC appear in the same order as 
the corresponding states  in the spin-1 case shown in Fig. \ref{fig1}. The only difference in the spin-2 case 
is the appearance of the triangular-lattice soliton which is absent in the spin-1 case. 
\begin{figure}[!t] 
\centering
\includegraphics[width=1\linewidth]{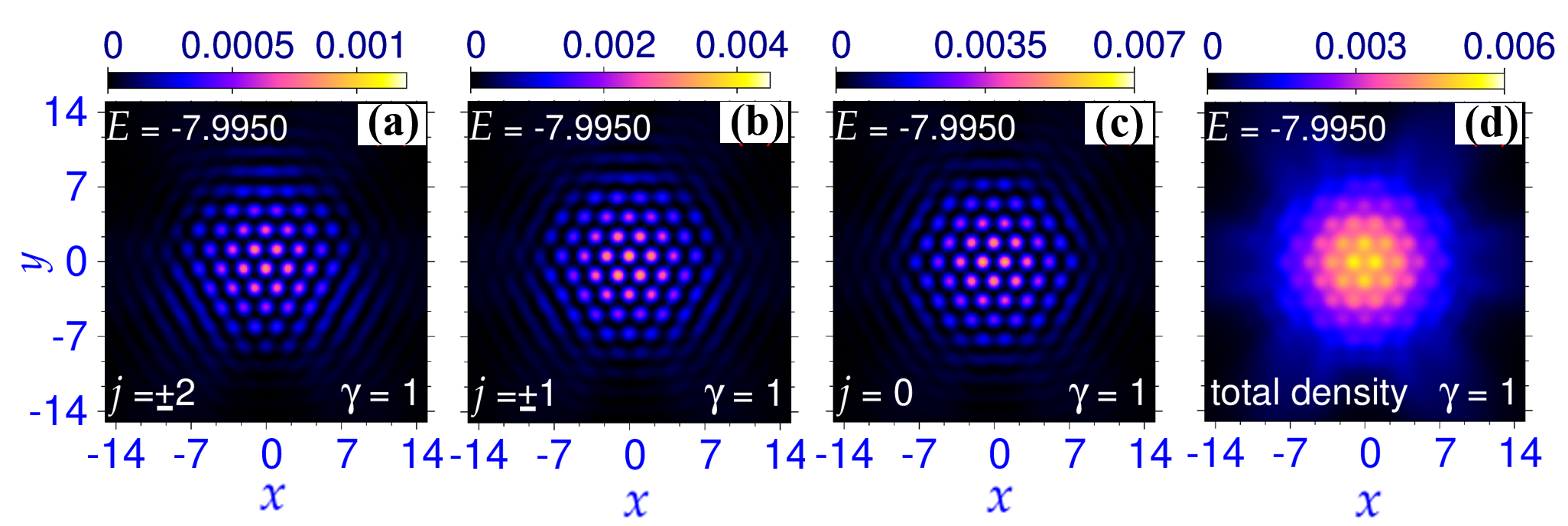} 
\caption{(a) Component density $n_{\pm 2}(\boldsymbol \rho)$, (b)  $n_{\pm 1}(\boldsymbol \rho)$,
(c) $n_0(\boldsymbol \rho)$, and (d) total density $n(\boldsymbol \rho)$ {of} a triangular-lattice soliton of
the SO-coupled spin-2 BEC with $c_0=-0.2$ and $\gamma=1$.
}
\label{Figure6&7}
\end{figure}  
We illustrate the formation of a $(-2,-1,0,+1,+2)$-type multiring  soliton for $c_0=-0.2$ and $\gamma$ = 4 having 
energy $-31.9966$ in Figs.~\ref{Figure5}(a)-(d), where we display the contour plots of (a) component densities $n_{\pm 2}(\boldsymbol \rho)$, 
(b) $n_{\pm 1}(\boldsymbol \rho)$, (c) $n_0(\boldsymbol \rho)$, and (d) total density $n(\boldsymbol \rho)$ of this soliton. 
The charge of phase singularities in $\psi_{\pm 2}$ and $\psi_{\pm 1}$ are ascertained from Figs.~\ref{Figure5}(e)-(h), 
where we display the contour plots of the phases of wave-function components of the 
$(-2,-1,0,+1,+2)$-type Rashba SO-coupled soliton. The phases correspond to phase changes of $\mp 4\pi$ and $\mp 2\pi$
in $j = \pm 2$ and $j=\pm 1$ components, respectively, under a complete rotation around the center. As discussed
earlier, we also have $(-1,0,+1,+2,+3)$-type multiring soliton that continues to be the solution of 
Eqs.~(\ref{cgpet3d-1})-(\ref{cgpet3d-3}) for large SO coupling strengths. In Fig.~\ref{Figure2&3}, 
we show its emergence for $c_0=-0.2$ and $\gamma=4$ via contour plots of densities  (a) $n_{+2}(\boldsymbol \rho)$, (b) $n_{+1}(\boldsymbol \rho)$, 
(c) $n_0(\boldsymbol \rho)$, (d) $n_{-1}(\boldsymbol \rho)$, and (e) $n_{-2}(\boldsymbol \rho)$  with energy $-31.9930$. Next in Fig.~\ref{Figure2&3}, 
we show the contour densities (f) $n_{+2}(\boldsymbol \rho)$ , (g) $n_{+1}(\boldsymbol \rho)$, (h) $n_{0}(\boldsymbol \rho)$, (i) $n_{-1}(\boldsymbol \rho)$, and (j) $n_{-1}(\boldsymbol \rho)$ of the 
asymmetric soliton for $c_0=-0.2$ and $\gamma=4$ and with energy $-31.9956$. For an intermediate 
SO-coupling strength of $\gamma = 1$, contour plots of (a) densities $n_{\pm 2}(\boldsymbol \rho)$, (b) $n_{\pm 1}(\boldsymbol \rho)$, (c) 
$n_{0}(\boldsymbol \rho)$  and (d) total density $n(\boldsymbol \rho)$ of the triangular-lattice soliton for $c_0=-0.2$ and with energy $-7.9950$ are shown in Figs.~\ref{Figure6&7}(a)-(d). This soliton has a hexagonal-lattice crystallization in component and total 
densities. Besides the asymmetric soliton, two other solitons, which appear explicitly for large SO-coupling 
strengths, are stripe and square-lattice solitons. In Fig.~\ref{Figure4&5} we present the contour plots of 
(a) densities $n_{\pm 2}(\boldsymbol \rho)$, (b) $n_{\pm 1}(\boldsymbol \rho)$, (c) $n_{0}(\boldsymbol \rho)$, and (d) total density $n(\boldsymbol \rho)$ of the stripe soliton of energy
$-31.9962$ for $c_0 =-0.2$ and $\gamma =4$. Similar to the stripe soliton in the SO-coupled spin-1 BEC, 
the component densities have a stripe pattern which is not present in their sum.
The scenario is different in the case of a square-lattice soliton. For example, the component and total densities of the square-lattice soliton of energy $-31.9950$ 
for $c_0 =-0.2$ and $\gamma =4$  are shown in  Fig.~\ref{Figure4&5}: (e) $n_{\pm 2}(\boldsymbol \rho)$, (f) $n_{\pm 1}(\boldsymbol \rho)$, (g) $n_{0}(\boldsymbol \rho)$, and (h) total 
density $n(\boldsymbol \rho)$,  where the square-lattice pattern in the total density as well as the component densities is explicit.  

\begin{figure}[!t] 
\centering
\includegraphics[width=1\linewidth]{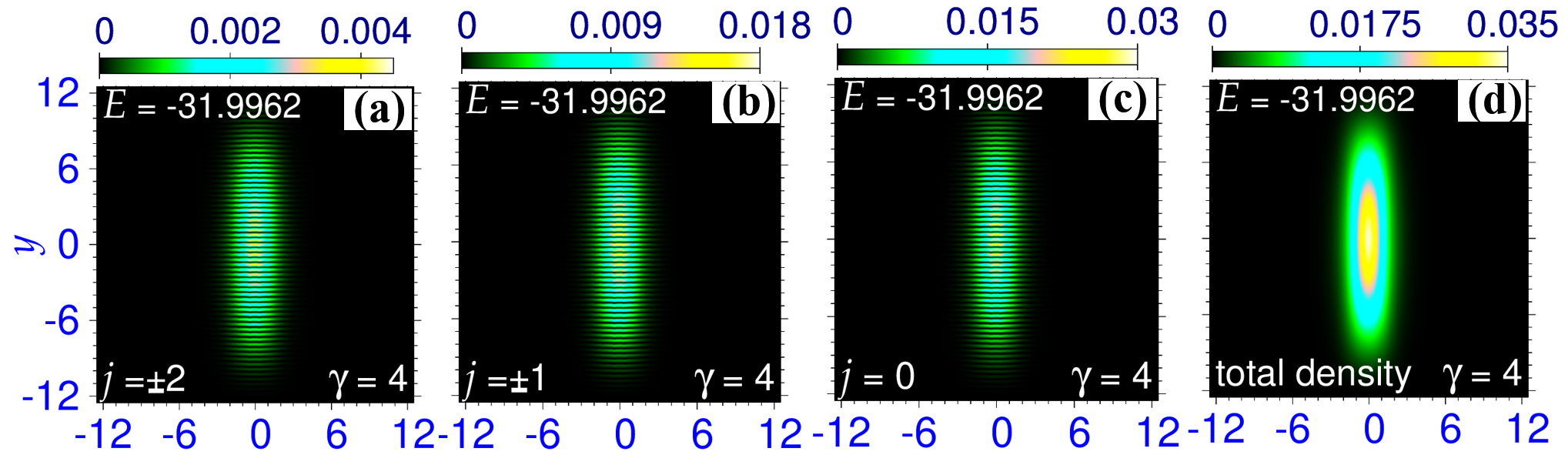} 
\includegraphics[width=1\linewidth]{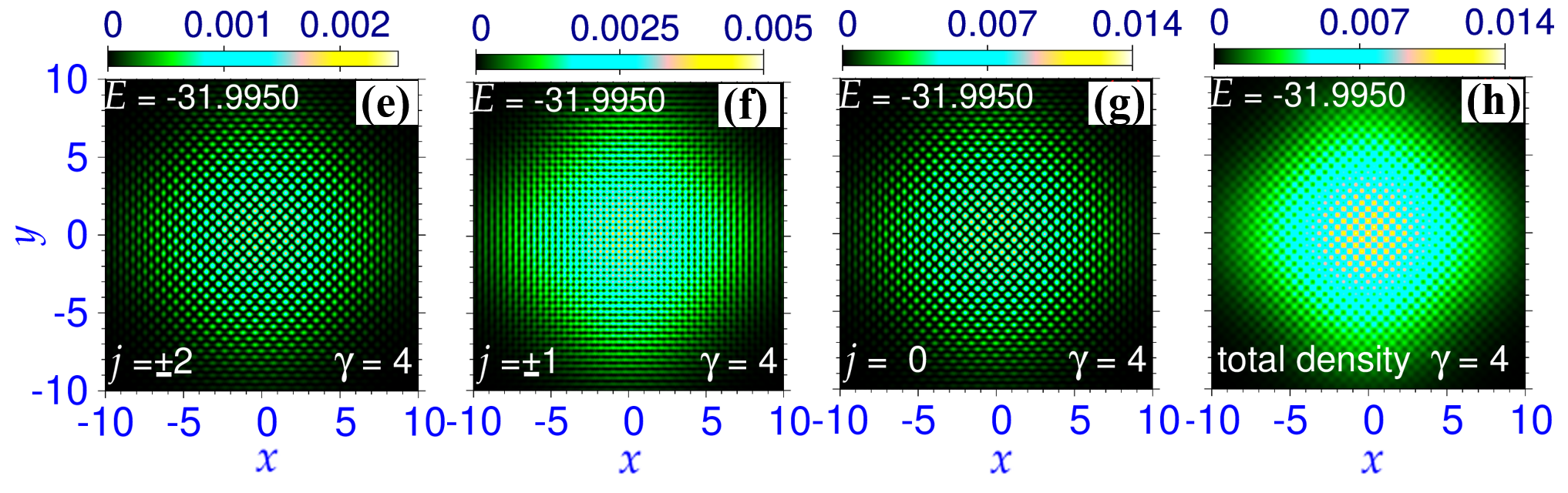} 
\caption{Upper row: (a) component density $n_{\pm 2}(\boldsymbol \rho)$, (b) $n_{\pm 1}(\boldsymbol \rho)$,  
(c) $n_0(\boldsymbol \rho)$, and (d) total density $n(\boldsymbol \rho)$ of a stripe Rashba SO-coupled 
soliton with $c_0=-0.2$ and $\gamma=4$. Lower row: (a) component density $n_{\pm 2}(\boldsymbol \rho)$, 
(b) $n_{\pm 1}(\boldsymbol \rho)$,  (c) $n_0(\boldsymbol \rho)$, and (d) total density $n(\boldsymbol \rho)$
of a square-lattice soliton with the same $c_0$ and $\gamma$.
}
\label{Figure4&5}
\end{figure}

\section{Summary of Results}
\label{section4}
We have investigated spontaneous spatial order in a Rashba SO-coupled nonmagnetic trapless quasi-2D spin-1 and
spin-2 BECs for various SO-coupling strengths $\gamma$ without spinor interactions
by numerically solving the coupled GP equations. For small 
SO coupling, $(−1, 0, +1)$ and $(0, +1, +2)$-type solitons are found for a spin-1 BEC, whereas $(-2,−1, 0, +1,+2)$ 
and $(-1, 0, +1, +2, +3)$-type solitons are found for a spin-2 BEC, which develop multiring structure with an 
increase in SO coupling strengths. For intermediate strengths of SO-coupling {($\gamma\approx 1$)}, 
in spin-2 BEC, triangular-lattice soliton having hexagonal crystallization in component as well as total density
appears as quasi-degenerate state. For relatively larger SO-coupling strengths, {asymmetric,} stripe 
and square-lattice solitons appear as quasi-degenerate solitons for spin-1 and spin-2 BECs. We also established 
the dynamical stability of these solitonic states by stable real-time simulation over a long interval of time as in the case of spin-1 and spin-2 ferromagnetic, anti-ferromagnetic and cyclic solitons \cite{SS-SPIN1,SS-spin-2} (result not shown here).  { The spatially-periodic structures are not possible in multicomponent BECs with only the contact interaction present.  The presence of the SO-coupling terms with first derivatives, mixing the components, are the necessary ingredients for the emergence of different types of spatially-periodic solitons.}
 In conclusion, it is strongly suggestive  that the spatially-periodic supersolid-like states found in SO-coupled pseudo spin-1/2 \cite{Ketterle,Li,SS-spin-1/2}, spin-1 \cite{SS-SPIN1}, and spin-2 \cite{SS-spin-2} spinor BECs are not a consequence of spinor interactions but are a consequence of multicomponent nature of these states in presence of SO coupling. 

\section*{Acknowledgments}
S.K.A. acknowledges support by the CNPq (Brazil) grant 301324/2019-0, and by the ICTP-SAIFR-FAPESP (Brazil) grant 
2016/01343-7. S.G. acknowledges support from the Science and Engineering Research Board, Department of Science and
Technology, Government of India through Project No. CRG/2021/002597.

\end{document}